**Pressure induced Superconductivity in Topological Compound $Bi_2Te_3$**

J. L. Zhang[1,2], S. J. Zhang[1], H. M. Weng[1], W. Zhang[1], L. X. Yang[1], Q. Q. Liu[1], S. M. Feng[1], X. C. Wang[1], R. C. Yu[1], L. Z. Cao[2], L. Wang[3], W. G. Yang[3], H. Z. Liu[4], W. Y. Zhao[5], S. C. Zhang[6], X. Dai[1], Z. Fang[1*], C. Q. Jin[1*]

[1] *Beijing National Laboratory for Condensed Matter Physics, and Institute of Physics, Chinese Academy of Sciences, Beijing 100190, China,*
[2] *Department of Physics, University of Science & Technology, Hefei, China*
[3] *HPsync, Geophysical Laboratory, Carnegie Institution of Washington, Advanced Photon Source, Argonne, IL 60439*
[4] *Natural Science Research Center, Harbin Institute of Technology, Harbin 150080, China*
[5] *State Key Laboratory of Advanced Technology for Materials Synthesis and Processing, Wuhan University of Technology, Wuhan 430070, China*
[6] *Department of Physics, McCullough Building, Stanford University, Stanford, CA 94305-4045, Center for Advanced Study, Tsinghua University, Beijing 100084, China*

$Bi2Te_3$ compound has been theoretically predicted (1) to be a topological insulator, and its topologically non-trivial surface state with a single Dirac cone has been observed in photoemission experiments (2). Here we report that superconductivity (T$c$ ~3K) can be induced in $Bi2Te_3$ as-grown single crystal (with hole-carriers) via pressure. The first-principles calculations show that the electronic structure under pressure remains to be topologically nontrivial, and the Dirac-type surface states can be well distinguished from bulk states at corresponding Fermi level. The proximity effect between superconducting bulk states and Dirac-type surface state could generate Majorana fermions on the surface. We also discuss the possibility that the bulk state could be a topological superconductor.

∗ Corresponding authors: E-mail: zfang@iphy.ac.cn & Jin@iphy.ac.cn

Discovery of the topological insulator (TI) state in HgTe/CdTe quantum wells (*3,4*), Bi1-*x*Sbx alloys (*5, 6*), and Bi2Te3, Bi2Se3, Sb2Te$_3$ crystals (*1, 2, 7*) has generated great interests in condensed matter physics. These materials have an insulating gap in the bulk, and gapless surface states which are protected by the time-reversal symmetry (*8,9*). Similar to TIs, time-reversal invariant topological superconductors have a full pairing gap in the bulk, and gapless Majorana states on the surface or edge (*10–13*). Majorana Fermions (*14*)—half of ordinary Dirac fermions, could be very useful in topological quantum computing (*15–17*). Whereas the topological superfluid state has been theoretically proposed for the B-phase of He3 (*10–13*), the search of topological superconductors in electronic systems remains to be a great challenge in condensed matter physics.

Majorana fermions could also arise at the interface between three dimensional TI and a conventional superconductor through proximity effect (*18*). However, if the TI side and the superconductor side are made from different materials, one may encounter serious mismatch problem at the interface. To overcome this problem, an ideal design is to make the doped TI superconducting and consider the proximity effect through the natural "interface" between the bulk and surface regions, without introducing any other superconducting compound. In other words, Majorana fermions may be realized on the surface of TI with the superconducting phase inside the bulk. Nevertheless, to reach such situation, two criteria should be satisfied: (1) superconductivity in the bulk states of TI; (2) still well defined Dirac type surface states, which can be well distinguished from the bulk states around the Fermi energy. In the usual case, introducing bulk carriers (which is necessary although not sufficient condition for the bulk superconductivity) will push the Fermi level into the bulk states, where the topologically protected surface states are ill defined. The combination of both conditions impose a stringent constraint, which may not be easily realized in the Cu*x*Bi2Se$_3$ superconductor where n type carriers are observed (*19*). We will show in this paper, that both superconductivity and well defined surface states can be realized simultaneously in p doped topological compound Bi2Te$_3$ via pressure.

The Bi2Te$_3$ single crystal was grown using flux method based on Bi2Te$_3$ stoichiometry (20), and the diamond anvil cell technique was used for the high pressure measurements. A superconducting phase with transition temperature T$c$ ~3K was observed in Bi2Te$_3$ as-grown single crystal in the pressure range from 3.1 to 6.3 GPa, where the same parent crystal structure remains. Fig. 1 shows the evolution of resistance as function of temperature of Bi2Te$_3$ single crystal at various pressures [see supporting online materials (SOM) for methods]. For pressure above 3.1 GPa, a clear superconducting transition was observed, and the transition temperature keeps almost a constant up to 6.3 GPa before a crystal structure phase transition occurs (see Fig. 3, discussed later). The inset of Fig. 1(a) shows the data collected at 5.0 GPa, where resistance drops to zero at low temperature, and the transition temperatures of onset, midpoint or zero resistance are defined based on the differential of resistance over temperature (dR/dT). The superconducting transition is sharp with the transition width (from 10% to 90% of the normal state resistance at $T^{onset}$) around 0.3K. This indicates the good homogeneity of superconducting phase. It is noted [see Fig. 1 (b)], although the high pressure phase (>6.3 GPa) of Bi2Te$_3$ is structurally different, it is also superconducting with higher $T^{onset}_c$ (>8K). Such superconducting high pressure phase (>6.3 GPa) has been reported before (21), and the different transition temperatures [above 8K in our measurement and around 2K reported in (21)] may originate from the different carrier type and density of the samples (as addressed below). The central result of our paper is the discovery of the superconducting phase between 3.1 GPa and 6.3 GPa, with the crystal structure same as in the ambient phase, where the TI behavior has been predicted and observed.

To assure what observed in Fig. 1 is indeed a superconducting transition we further conducted the measurements around the transition temperature at variant external magnetic field. Fig. 2(a) shows the measured resistance at 6.1 GPa with applied magnetic *H*. It is obvious that the transition temperature Tc decreases with increasing magnetic field. This shows strong evidence that the transition is superconductivity in nature. The inset of Fig. 2(a) shows the change of Tc with magnetic field $H$. Using the Werthdamer-Helfand-Hohenberg formula (*22*) of $Hc2(0) = -0.691[dH_{c2}(T)/dT]_{T=T_c} \cdot T_c$, the upper critical field $Hc2(0)$ is extrapolated to be 1.83 T for $H_{//}$ c (here the single crystal is placed inside the diamond anvil cell with magnetic H direction perpendicular to *ab*-plane).

The Hall coefficient measurements indicated the p type carrier nature of the Bi2Te$_3$ as-grown single crystal [as shown in Fig. 2(b)]. The carrier density calculated from the linear part at high magnetic field H is approximately $3\sim6\times10^{18}/cm^3$ as deduced for several measurements at ambient pressure. The carrier density changes slightly from high temperature to low temperature as show in the inset of Fig. 2(b). Both p and n type carriers have been reported for Bi2Te$_3$ as-grown single crystal depending on the growth conditions and the elements ration of Bi over Te in the starting materials. In our samples, slightly excess Te helps to form p type compound. This is different from Cu$x$Bi2Se$_3$ superconductor where n type carriers are observed (*19*). Direct Hall measurement at high pressure is difficult, however primary measurements indicate little change in the carrier density from ambient to 1.0 GPa (see SOM Fig. S1). In addition, the relative change of carrier density with pressure can be estimated from the transport data. The resistivity at fixed temperature (10K) is only changed by a factor of 3 from ambient pressure to 3.1 GPa, where superconductivity occurs. Therefore, we do not expect the dramatic change of carrier density at the superconducting pressure.

To confirm that the superconducting phase transition is not caused by crystal structure phase transition under pressure, we further conducted synchrotron X-ray diffraction experiments with pressure above 30 GPa in an average 2 GPa step at room temperature. Fig. 3 shows the X-ray diffraction patterns of Bi2Te$_3$ polycrystalline (ground from the single crystal) at several selected pressures. The ambient phase remains stable up to 6.3 GPa [the resistance measurements indicate an intermediate phase occurred at 6.8 GPa as shown in Fig. 1(b)], above which a structure phase transition was observed as seen in Fig. 3 for X-ray data under 8.6 GPa consistent with previous reports (*21*). Therefore the superconductivity observed in our experiments is surely from the parent structure of Bi2Te$_3$ that preserves the topological property as we will show in the electronic structure calculation. We further refined the crystal structure at high pressure using Rietveld method with GSAS package, and determined the lattice parameters and atomic positions (see SOM Fig. S2).

Stoichiometric $Bi_2Te_3$ has layered hexagonal structure with unit cell consisting of five atomic layers, which form so-called quintuple-layer (QL). Its topological property has been theoretically predicted from first-principles calculations and analytical models (*1*), and confirmed by the ARPES measurement of surface states (*2*, *23*). Due to the strong spin orbit coupling, the band structures around the $\Gamma$ point of Brillouin zone (BZ) are inverted, similar to the case of HgTe quantum wells (*3*, *4*), which gives rise to the TI behavior.

By the first-principles calculations based on density functional theory and the generalized gradient approximation, we calculated the electronic structures of $Bi_2Te_3$ at both 0.0 GPa and 4.0 GPa, using the experimental lattice parameters and atomic positions (obtained from the X-ray refinement, see SOM Table S1). For the calculations of surface states, we use a thick slab consisting of 40 QLs. As shown in Fig. 4 (a) and (b), for the 0.0 GPa phase, our calculations well reproduce the earlier results (*1*, *23*). In particular, the calculated surface states can be well compared to the ARPES measurement (*2*, *23*). Given the small changes of lattice parameters and atomic positions at 4.0 GPa, we would not expect dramatic change of its electronic structure. Indeed, as shown in the Fig. 4 (c) and (d), the electronic structure at 4.0 GPa is quite similar to that at 0.0 GPa. Most importantly, by calculating the $Z_2$ number from parity analysis and the surface states, we confirm that the compound remains to be topologically non-trivial at 4.0 GPa.

$Bi_2Te_3$ has the particular property that the surface states are located around $\Gamma$ point, while both conduction band minimum (CBM) and valence band maximum (VBM) of bulk states are located away from the $\Gamma$ point. In other words, for the low energy range, the surface states and bulk states are well separated in the momentum space (see Fig. 4). The direct band gap at $\Gamma$ point is much bigger than the indirect bulk band gap, which is formed from states around the Z point of the bulk BZ. Such character is crucial for the criteria discussed at beginning to realize the bulk-to-surface proximity effect. Using the rigid band model, we can estimate the Fermi level position as functions of carrier concentration, and find that the surface states remain to be well defined even if the carrier density is as high as $10^{21}/cm^3$ (hole type). For our experimental situation, where superconductivity was observed, the carrier density (of the order $10^{19}/cm^3$) is much smaller than this level. The electronic structures at 4.0 GPa are in fact quantitatively different with that of 0.0 GPa in the following senses: (1) the direct gap at $\Gamma$ point is enhanced while the indirect bulk band gap is reduced; (2) the separation between surface states and bulk states at the Fermi level is further enhanced. We have also analyzed the penetration depth of surface states around the Fermi energy determined by the experiments [k1 and k2 points shown in Fig. 4(d)], and find that they are exponentially localized to the surface region with half-width about 3 QLs. In contrast to Bi2Te3, for the electron-doped Bi2Se3, the CBM of bulk states locates around the $\Gamma$ point, and very close to the surface states. Both facts support our conclusion that the topological surface states can maintain their character in the presence of the p type bulk carriers, and the resulting proximity effect with the bulk superconducting carriers can give rise to Majorana fermions in the surface state.

We now turn to the exciting possibility that the bulk superconducting state in Bi2Te$_3$ could be a topological superconductor (*10–13, 24–27*). In insets of Fig. 4(a) and (c), we see that with p type doping, holes form disconnected pockets. Pairing amplitude on a given hole pocket could be largely determined by the phonon contribution mediated at the small momentum transfer, and a uniform pairing order parameter could be established on each hole pocket. The relative pairing amplitude among the different hole pockets would be determined by the large momentum transfer, where the Coulomb repulsion plays a more dominant role. Such a repulsive interaction generally favors opposite pairing amplitudes on different fermi pockets, leading to a negative Josephson coupling among the neighboring hole pockets. However, due to the three-fold symmetry of the Fermi surface, such a coupling is frustrated, so that the pairing order parameter in the ground state may become complex. A natural choice of such a complex orbital pairing symmetry without breaking the time reversal symmetry is a triplet pairing symmetry similar to the BW state in He3-B-phase. Such a pairing state has been shown on general grounds to be a topological superconducting state respecting the time reversal symmetry (*10–13,24–27*).

In summary, we have experimentally observed superconductivity in $Bi_2Te_3$ under pressure, where the crystal structure remains the same as the ambient topological phase. Topological surface states remain to be well defined under pressure, and in the presence of bulk p type carriers, and this natural bulk-surface proximity effect could give rise to Majorana fermions on the surface. We also discussed the pairing symmetry of the bulk superconducting state, and addressed the possibility that it could realize the 3D topological superconductor with time reversal symmetry. We plan to experimentally investigate the pairing symmetry in the bulk and to detect the Majorana fermion on the surface directly. We also plan to induce the superconducting state in Bi2Te3 using strain, by growing thin films on substrate with well designed lattice mismatch.

ACKNOWLEDGMENTS. We thank the Natural Science Foundation (NSF) and Ministry of Science and Technology (MOST) of China through the research projects (10820101049, 2007CB925000, and 2010CB923000), and the International Science and Technology Cooperation Program of China. HPCAT is supported by CIW, CDAC, UNLV and LLNL through funding from DOE-NNSA, DOE-BES and NSF. HPSynC is supported as part of EFree, an Energy Frontier Research Center funded by DOE-BES under Award DE-SC0001057. S.C.Z. is supported by the NSF under Grant DMR-0904264.

28 We acknowledge the supports from the NSF of China (10820101049), the 973 Program of China (No. 2007CB925000 and 2010CB923000), and the International Science and Technology Cooperation Program of China.

**Figure Captions;**

Figure 1: (a) (Upper panel) The resistance of Bi2Te3 single crystal as function of temperature at various pressure showing a superconducting transition around 3K. The inset is the resistance near the critical temperature Tc with the differentiation of dR/dT showing the definition of $Tc^{onset}$, $Tc^{mid}$ and $Tc^{zero}$, respectively; (b) (lower panel) The enlarged (low temperature) part of the measured resistance, indicating the almost constant Tc 3K for the states between 3.1 GPa and 5.0 GPa, and a much higher Tc of 8.1 K at 8.9 GPa (which is structurally different). A broad superconducting transition is observed for the intermediated pressure range, such as the data shown for 6.8 GPa.

Figure 2: (a) (Upper panel) The superconducting transition of Bi2Te3 single crystal at 6.1 GPa with applied magnetic field H perpendicular to the *ab*-plane of single crystal. Inset shows the change of Tc with *H*. (b) (Lower panel) The Hall resistance of Bi2Te3 single crystal with magnetic field H perpendicular to the *ab*-plane of the single crystal. Inset is the carrier density deduced from the linear part of Hall resistance at various temperatures at ambient pressure.

Figure 3: The synchrotron X-ray diffraction patterns of Bi2Te3 samples at selected pressures showing the ambient structure is stable at least up to 6.3 GPa, and a structurally different phase appears at 8.6 GPa (the circle indicating the peaks from new structure).

Figure 4: The calculated bulk electronic structures (left panels) and surface states (right panels) of the Bi2Te3 at 0.0 GPa [(a) and (b)] and 4.0 GPa [(c) and (d)], respectively. The chemical potential corresponding to the experimental hole concentration ($5.8 \times 10^{18}$/cm$^3$) are indicated as horizontal dashed lines and the corresponding Fermi surfaces are also shown as insets. In order to understand the k-space separation of bulk and surface states, we show the surface states schematically in the bulk band structures (left panels).

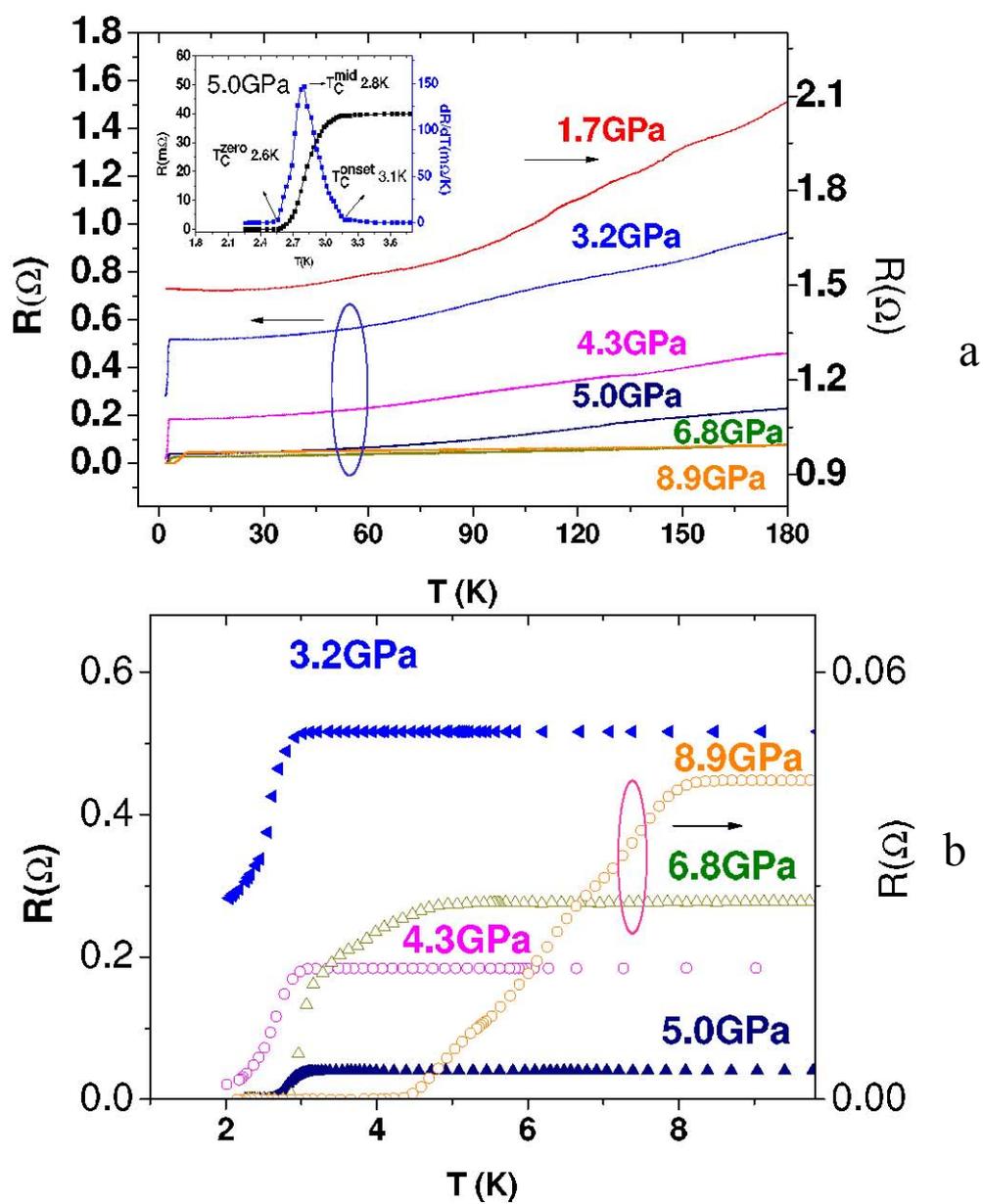

Fig. 1

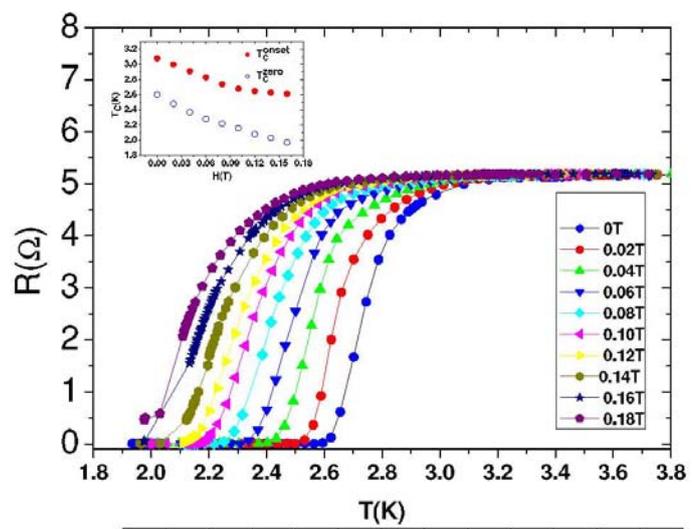

a

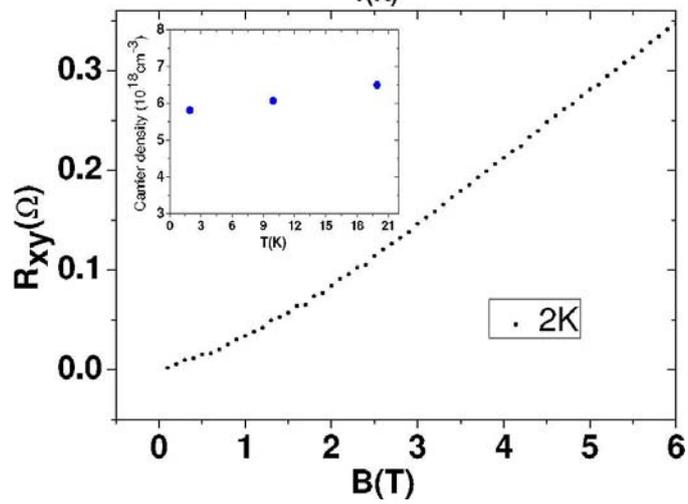

b

Fig. 2

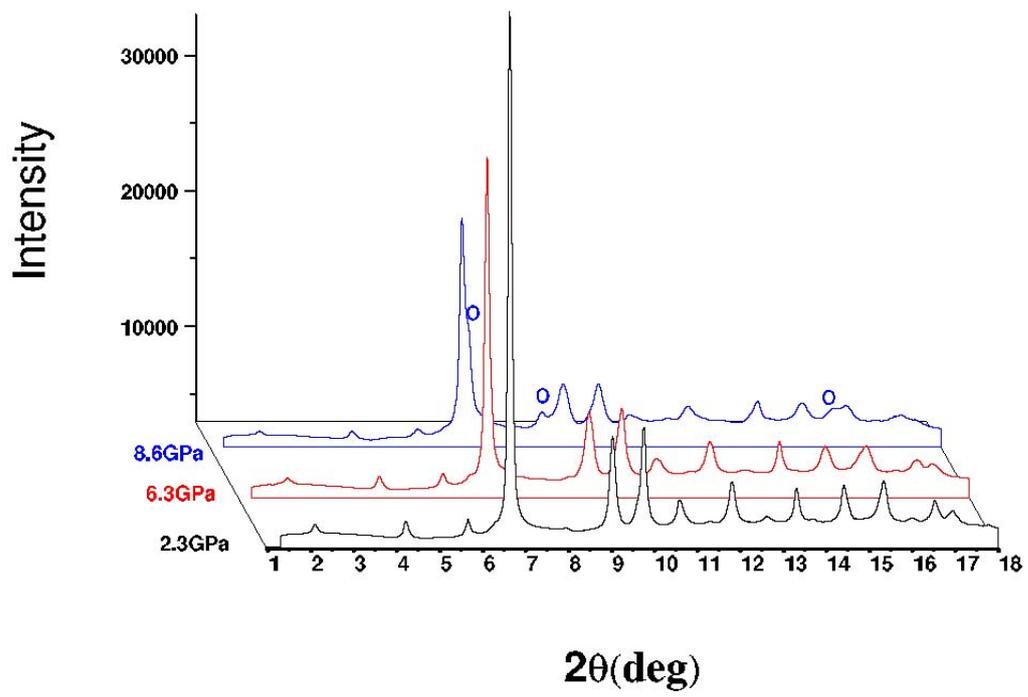

Fig.3

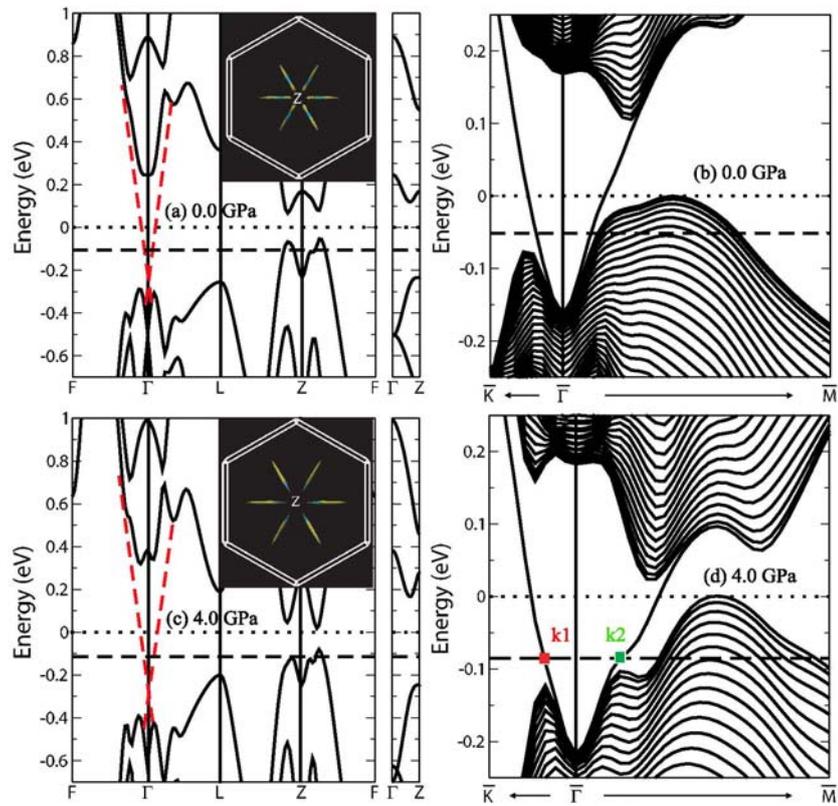

Fig. 4

# Supporting Information

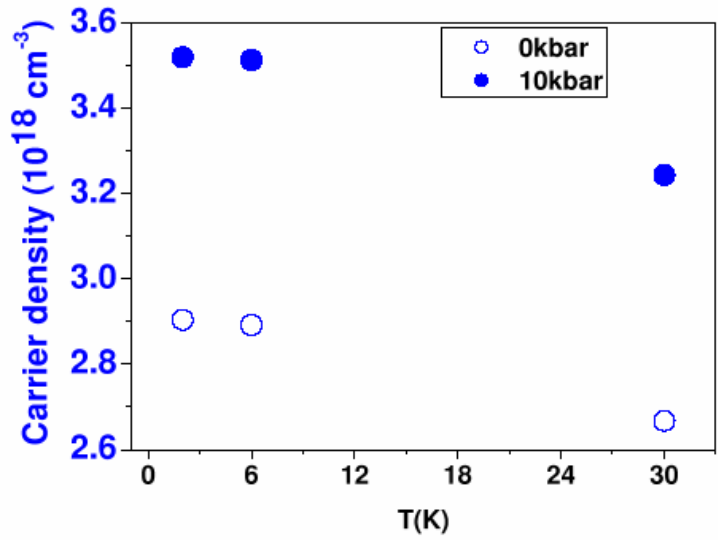

Figure S1: The carrier density of Bi$_2$Te$_3$ single crystal at ambient and 1.0 GPa, respectively, showing a slight increase with pressure.

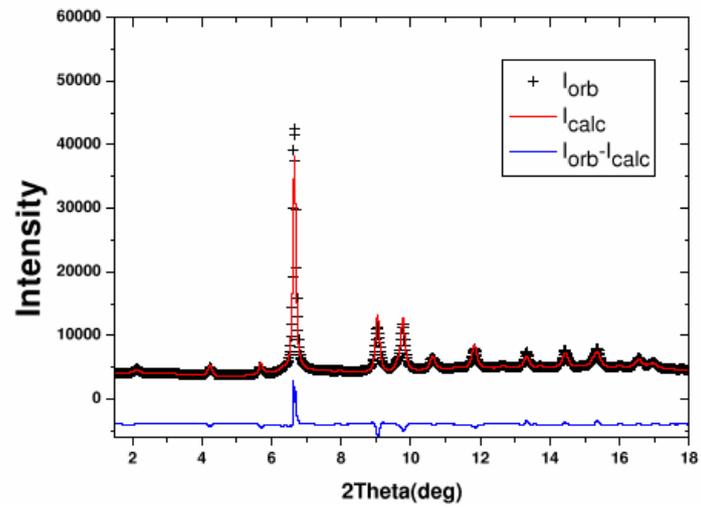

Figure S2: The Rietveld refinements of the crystal structure for Bi$_2$Te$_3$ at 4.0 GPa at room temperature. The cross and red line correspond to that for experiment and refinement, respectively, while the blue is the difference.

Table S1: The crystal structure data of Bi$_2$Te$_3$ at ambient and 4.0 GPa at room temperature. The theoretically relaxed lattice constants and internal atomic sites are listed in parenthesis.

| | atom | site | x | y | z | | |
|---|---|---|---|---|---|---|---|
| 0GPa | Bi | 6c | 0 | 0 | 0.400 (0.400) | | |
| | Te1 | 3a | 0 | 0 | 0.0 (0.0) | | |
| | Te2 | 6c | 0 | 0 | 0.2095 (0.2094) | | |
| | Bi-Te1 (Å) | Bi-Te2 (Å) | Te1-Bi-Te1 ° | Te2-Bi-Te2 ° | d1 (Å) | d2 (Å) | d3 (Å) |
| | 3.2551 | 3.0808 | 84.805 | 90.874 | 2.0434 | 1.7515 | 2.6243 |
| | a = b = 4.390 (4.400) Å, c = 30.497 (30.335) Å, α = 90°, β = 90°, γ = 120° | | | | | | |
| 4GPa | atom | site | x | y | z | | |
| | Bi | 6c | 0 | 0 | 0.4030 (0.401) | | |
| | Te1 | 3a | 0 | 0 | 0.0 (0.0) | | |
| | Te2 | 6c | 0 | 0 | 0.2059 (0.2055) | | |
| | Bi-Te1 (Å) | Bi-Te2 (Å) | Te1-Bi-Te1 ° | Te2-Bi-Te2 ° | d1 (Å) | d2 (Å) | d3 (Å) |
| | 3.1845 | 2.9731 | 83.331 | 90.802 | 2.0411 | 1.6924 | 2.2989 |
| | a = b = 4.234 (4.237) Å, c = 29.296 (29.262) Å, α = 90°, β = 90°, γ = 120° | | | | | | |

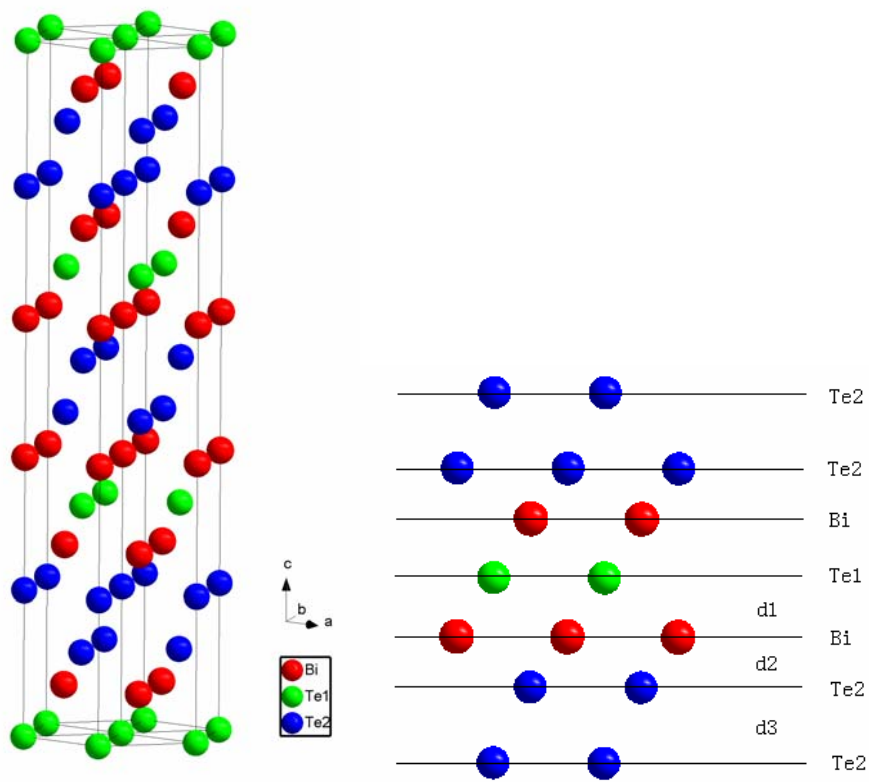

Figure S3: The crystal structures of Bi$_2$Te$_3$.